\documentclass{kapproc} 

\usepackage{procps} 

\usepackage[dvips]{graphicx}

\upperandlowercase

\kluwerbib 

\newcommand{\lya}{Ly$\alpha$}
\newcommand{\ha}{H$\alpha$}
\newcommand{\nii}{[N{\sc II}]}
\newcommand{\kms}{\hbox{km\,s$^{-1}$}}

\begin{document}

\articletitle[Proto-clusters associated with radio galaxies 
from {\it z} = 2 to {\it z} = 4]
{Proto-clusters associated with\\ 
radio galaxies from {\it z} = 2 to {\it z} = 4}

\author{Jaron Kurk,\altaffilmark{1,2} Bram Venemans,\altaffilmark{1}
Huub Röttgering,\altaffilmark{1} George Miley\altaffilmark{1} and\\
Laura Pentericci\altaffilmark{3}}

\altaffiltext{1}{Sterrewacht Leiden\\
P.O. Box 9513, 2300 RA, Leiden, The Netherlands}
\altaffiltext{2}{INAF, Osservatorio Astrofisico di Arcetri\\
Largo Enrico Fermi 5, 50125, Firenze, Italy}
\altaffiltext{3}{Max-Planck-Institut für Astronomie\\
Königstuhl 17, D-69117, Heidelberg, Germany}
\email{kurk@arcetri.astro.it}

\anxx{Kurk\, Jaron} 

\begin{abstract}

We have carried out narrow band imaging targeted at the \lya\ line of
emitters associated with radio galaxies at $2 < z < 5$. Subsequent
spectroscopy has confirmed the identity of $> 120$ \lya\ emitters and
led to the discovery of six proto-clusters up to $z = 4.1$ with
velocity dispersions between 300 and 1000 \kms. The number of emission
line galaxies in the observed fields is four to fifteen times higher
than in blank fields. These results strongly support the idea that
high redshift radio galaxies are the progenitors of central brightest
cluster galaxies located in the progenitors of clusters of galaxies.

\end{abstract}


\section{Introduction}
Observational studies of cosmological structures can never be complete
without the full picture provided by multi wavelength explorations,
but in the case of the search for very high redshift clusters or their
progenitors the use of multi wavelength observations is
particularly fruitful. The detection of galaxy aggregations using
conventional optical and X-ray becomes difficult at $z > 1$ due to the
presence of many fore- and background objects and/or the surface
brightness dimming of the extended emission. Several lines of
observation indicate that powerful radio galaxies, as are found up to
$z = 5.2$, are located within regions of high galaxy density. The
technique of narrow band imaging can detect a group of emission line
galaxies at a narrow high redshift range. Applying this technique to
fields containing high redshift galaxies, we succeed to find galaxy
groups at high redshift. The groups will be used to study cluster
progenitors, its galaxies and the massive radio galaxy hosts.

\section{Sample and observations}
We had initially selected ten high redshift radio galaxies (HzRGs, $z
> 2$) from our compendium of about 150 $z > 2$ radio sources, many of
which have been found in the Leiden survey of steep spectrum radio
sources (\cite{rot94,bre00}). We have succesfully observed six fields
and during an extension of the program, also the field of the most
distant known radio galaxy, at $z = 5.19$, bringing the total number
of nights allocated to this VLT Large Program to twenty. The observing
time was more or less evenly distributed over imaging and the
subsequent multi object spectroscopy, both carried out with the FORS2
instrument. The detector of this instrument has a field of view of
almost 7 $\times$ 7 arcminutes (about 3 $\times$ 3 Mpc at $z > 2$).
For the two highest redshift sources custom narrow band filter were
manufactured. For the others we have used available FORS filters. In
the following, two observed fields will be described in more detail
and a summary of the results of the program will be presented.

\section{Early results I: MRC 1138$-$262 at $z = 2.16$}
This radio galaxy was the target of our pilot study carried out with
FORS1 at the VLT. It is exceptionally suitable as it manifests many of
the properties which indicate a rich environment, namely the highest
radio rotation measure in a sample of 80 HzRGs and a very distorted
radio morphology (\cite{car97}), a very clumpy UV morphology
(\cite{pen97}), a large and luminous \lya\ halo ($\sim$ 150 kpc,
\cite{kur00b}), and extended X-ray emission (\cite{car02}). In
addition, its redshift of 2.16 is near the lower limit where detection
of \lya\ emission with ground based facilities is possible.  Narrow
band imaging of the field of 1138$-$262 resulted in the detection of
50 candidate \lya\ emitters (LAEs, \cite{kur00}). Multi object
spectroscopy of these candidates confirms the presence of a single
emission line at the expected wavelength for fifteen objects
(\cite{pen00a}). One of these is an AGN as shown by the broadness of
its emission line (FWHM $\sim$ 6000 \kms). The others seem to form two
dynamical groups with velocity dispersions of $\sim$ 200 and 400 \kms\
(if regarded as one group the dispersion is $\sim$ 1000 \kms). The
volume density of LAEs is about four times higher than that measured
for the blank field population of LAEs (\cite{ste00}). In addition to
the observations necessary to identify the LAEs, we have obtained
imaging in a number of optical and infrared bands with FORS and ISAAC
at the VLT, one of which is a narrow band filter in the NIR targeted
at redshifted \ha\ emission of galaxies associated with
1138$-$262. Excess narrow band flux testifies the presence of 40
candidate \ha\ emitters in the field (\cite{kur03a}). Long slit ISAAC
spectroscopy of nine candidates shows that three objects exhibit
emission lines identified as \ha\ and \nii\ while the remaining six
objects have fainter single lines consistent with the identification
with \ha, one of which has a FWHM of $\sim$ 5000 \kms\ revealing
another AGN at $z = 2.16$. The two AGN discovered with the narrow band
technique are part of a larger group of five AGN probably associated
with the radio galaxy as shown by Chandra observations of the field
(\cite{pen02}). The number density of soft X-ray sources detected by
Chandra is in excess of that found in other fields. The detected
overdensities of \lya, \ha\ and X-ray emitters draw a consistent
picture of a powerful radio galaxy in an aggregation of galaxies which
may form a cluster of galaxies.

\section{Early results II: TN J1338$-$1942 at $z = 4.11$}

With regard to both its continuum and \lya\ emission, TN J1338$-$1942
is amongst the most luminous radio galaxies known.  Its \lya\ line
profile and radio morphology are very asymmetric indicating strong
interaction with dense gas ({\cite{bre99}). Based on the line
equivalent width derived from the narrow and broad ($R$) band images
and also on the absence of emission in the broad $B$ band, 28 objects
in the field of 1338$-$1942 were selected as candidate LAEs
(\cite{ven02}). Of the 23 candidates subsequently observed
spectroscopically, 20 show a single emission line in the expected
wavelength range. The identification of the observed features with
lines other than \lya\ can be excluded in practically all cases. The
velocity distribution of the emitters has a dispersion of 326 \kms,
significantly smaller than expected from a random distribution of
emitters selected with the narrow band filter. Compared with the LALA
survey (\cite{rho00}), the volume density of LAEs is about fifteen
times larger than in a blank field. The spatial distribution of
emitters is not homogeneous over the observed field. The radio galaxy
is not in the center of the observed structure but rather at the
northern edge.

\section{High redshift proto-clusters and radio galaxies}

At the time this talk was delivered, six HzRG fields were observed
sufficiently deep to assess the density of companion LAEs.  In the
fields of HzRGs at $z$ = 2.86, 2.92, 3.13 and 3.14 the number of
candidate (and spectroscopically confirmed) emitters are: 52 (37), 70
(30), 78 (31) and 20 (11), which amounts to over-densities of 4 to
15. The velocity dispersion of the ensembles of emitters observed
decreases with increasing redshift: from $\sim$ 1000 \kms\ at $z = 2$
to $\sim$ 325 \kms\ at $z = 4$. It would be premature to draw
conclusions on this result, however, as there are only six data
points. The size of the structures of LAEs is in all cases larger than
the field sampled: $>$ 3 Mpc (comoving) and seems sometimes bound in
one direction. We have compared the number of spikes in the redshift
distribution of Lyman Break Galaxies (LBGs) at $2.7 < z < 3.4$
(\cite{ste98}) with the number of powerful (P$_{\rm 2.7\,GHz} >
10^{33}$ erg\,s$^{-1}$\,Hz$^{-1}$\,sr$^{-1}$) radio galaxies in this
redshift range (\cite{dun90}). Assuming that the radio sources are
active only once for a period of $\sim 10^7$ year, the numbers are
consistent with every LBG redshift spike being associated with a
massive galaxy that has been or will become a luminous radio source
once.

From the preliminary results we derive the following conclusions:
narrow band imaging is an efficient technique to find galaxy
over-densities in a narrow redshift range, HzRGs are excellent tracers
of these over-densities and may be the progenitors of central
brightest cluster galaxies located in the progenitors of clusters of
galaxies.

\chapbblname{kurk_myk03}
\chapbibliography{kurk_myk03.bib}

\end{document}